\documentclass[pre,aps,preprint,showpacs,preprintnumbers,amsmath,amssymb,secnumroman,eqsecnum]{revtex4}

\usepackage{graphicx}
\usepackage{dcolumn}
\usepackage{bm}

\newcommand{\beq}{\begin{equation}}
\newcommand{\eeq}{\end{equation}}
\newcommand{\beqa}{\begin{eqnarray}}
\newcommand{\eeqa}{\end{eqnarray}}
\newcommand{\vc}[1]{\mbox{\boldmath $#1$}}

\newcommand{\vol}[1]{{\bf #1}}

\newcommand{\du}[1]{{\bf\sf #1}}

\begin{document}


\title{Stokesian swimming of a sphere by radial helical surface wave}

\author{B. U. Felderhof}

 \email{ufelder@physik.rwth-aachen.de}
\affiliation{Institut f\"ur Theorie der Statistischen Physik\\ RWTH Aachen University\\
Templergraben 55\\52056 Aachen\\ Germany\\
}%



\date{\today}

\begin{abstract}
The swimming of a sphere by means of radial helical surface waves is studied on the basis of the Stokes equations. Explicit expressions are derived for the matrices characterizing the mean translational and rotational swimming velocities and the mean rate of dissipation to second order in the wave amplitude.
\end{abstract}

\pacs{47.15.G-, 47.63.mf, 47.63.Gd, 87.17.Jj}
\maketitle
\section{\label{I}Introduction}
Swimming microorganisms often rotate steadily about the swimming direction in a screw-type motion. The mean rotational swimming velocity can in principle be calculated from the bilinear theory of swimming, but there have been few investigations of the effect. In the following we study Stokesian swimming of a sphere by means of a radial helical surface wave.

In the studies of a swimming sphere by Lighthill \cite{1}, Blake \cite{2}, and Felderhof and Jones \cite{3} the stroke was assumed to be axial, so that the rotational swimming velocity vanishes. A particular example of rotational swimming of a sphere was presented in the context of potential flow \cite{4}. Pedley et al. have extended the work of Lighthill and Blake to include axisymmetric azimuthal flow \cite{5}. This leads to a nonvanishing mean rotational swimming velocity. Recently we have studied swimming of a sphere by tangential helical surface wave \cite{6}. The restriction to tangential flow was made on the assumption that it is a consequence of the structure of the swimmer. It has been argued that the restriction is plausible for cyanobacteria \cite{7},\cite{8}.

Swimming by means of a transverse surface wave was first studied by Taylor for a planar sheet \cite{9}. The corresponding effect for a sphere was investigated by Lighthill \cite{1}, and Blake \cite{2}, but with the restriction to axial stroke. Taylor also studied the swimming of a cylindrical body \cite{10} and found that this in general leads to a rotational swimming velocity. The swimming by means of a transverse helical wave on a finite body was studied by Ehlers and Koiller \cite{11}, who argued that the model is plausible for the cyanobacterium $Synechococcus$. They took the rest shape to be a spheroid and employed a tangential plane approximation method based on Taylor's work \cite{9}. We show that for a sphere the problem can be solved exactly.

For given radial surface wave we can find the mean translational and rotational swimming velocities and the mean rate of dissipation. The solution is based on an expansion of the surface displacement in terms of solutions of the Stokes equations chosen such that the displacement is radial at the undisplaced spherical surface. This choice of basis leads to a compact matrix formulation of the problem. As usual it is of interest to maximize the mean translational swimming velocity for given power \cite{12},\cite{13}. The optimization leads to a generalized eigenvalue problem which can be solved analytically for 2-mode swimmers. For many modes the eigenvalue problem can be solved numerically.

\section{\label{II}Flow equations}

We consider a sphere of radius $a$ immersed in a viscous
incompressible fluid of shear viscosity $\eta$.
The fluid is set in motion by distortions of the
spherical surface which are periodic in time and lead to a time-dependent flow field and swimming
motion of the sphere. We consider in particular radial helical distortions corresponding to a set of solutions of the Stokes equations,
\begin{equation}
\label{2.1}\eta\nabla^2\vc{v}-\nabla p=0,\qquad\nabla\cdot\vc{v}=0.
\end{equation}
 of the form
 \begin{eqnarray}
\label{2.2}
\vc{q}_{lm}(\vc{r})&=&\bigg(\frac{2l+2}{l(2l+1)}\hat{\vc{A}}_{lm}-\frac{2l-1}{2l+1}\hat{\vc{B}}_{lm}\bigg)\bigg(\frac{a}{r}\bigg)^l+\frac{l-2}{l}\hat{\vc{B}}_{lm}\bigg(\frac{a}{r}\bigg)^{l+2},\nonumber\\ p_{lm}(\vc{r})&=&2\eta(2l-1)(-1)^mP^m_l(\cos\theta)e^{im\varphi}\frac{a^l}{r^{l+1}},
\end{eqnarray}
in spherical coordinates $(r,\theta,\varphi)$, vector spherical harmonics $\hat{\vc{A}}_{lm},\hat{\vc{B}}_{lm}$ in the notation of Ref. 14 (with $2^{l+1}$ in the normalization coefficient replaced by $2l+1$), and with associated Legendre functions $P^m_l$ in the notation of Edmonds \cite{15}. It may be checked that at the surface $r=a$ the vector $\vc{q}_{lm}$ is in the radial direction. The radial component at the surface is
\begin{equation}
\label{2.3}q_{lm,r}(a,\theta,\varphi)=\frac{2l+2}{l}\;(-1)^mP^m_l(\cos\theta)e^{im\varphi},
\end{equation}
proportional to the local pressure disturbance.

A linear superposition of modes with coefficients oscillating at frequency $\omega$, corresponding to a first order velocity $\vc{v}_1(\vc{r},t)=\vc{v}_\omega(\vc{r})\exp(-i\omega t)$, leads to a mean translational swimming velocity $\overline{\vc{U}}_2=\overline{U}_2\vc{e}_z$ to second order in the surface displacements, as well as to a mean rotational swimming velocity $\overline{\vc{\Omega}}_2=\overline{\Omega}_2\vc{e}_z$. The swimming velocities are calculated from \cite{16}
\begin{equation}
\label{2.4}\overline{U_2}=-\frac{1}{4\pi}\int\overline{\vc{u}}_{2S}\cdot\vc{e}_z\;d\Omega,\qquad\overline{\Omega_2}=-\frac{3}{8\pi a}\int(\vc{e}_r\times\overline{\vc{u}}_{2S})\cdot\vc{e}_z\;d\Omega,
\end{equation}
where the integral is over spherical angles $(\theta,\varphi)$ and $\overline{\vc{u}}_{2S}$ is the mean second order surface velocity given by
\begin{equation}
\label{2.5}\overline{\vc{u}}_{2S}(\vc{s})=-\frac{1}{2}\mathrm{Re}(\vc{\xi}^*_\omega\cdot\nabla)\vc{v}_\omega\big|_{r=a},
\end{equation}
calculated from the first order velocity and the corresponding displacement vector $\vc{\xi}(\theta,\varphi,t)$, whose time derivative equals $\vc{v}_1$ at the surface $r=a$. The second order rate of dissipation may be
expressed as a surface integral \cite{4}
\begin{equation}
\label{2.6}\mathcal{D}_2=-\int_{r=a}\vc{v}_{1}\cdot\vc{\sigma}_{1}\cdot\vc{e}_r\;dS,
\end{equation}
where $\vc{\sigma}_1$ is the first order stress tensor. The mean rate of dissipation equals the power
necessary to generate the motion.

We consider first 2-mode linear superpositions of the form
 \begin{eqnarray}
\label{2.7}\vc{v}^c_1(\vc{r},t)&=&-\omega a\big[\gamma_{lm}\vc{q}_{lm}(\vc{r})+\gamma_{l+1,m}\vc{q}_{l+1,m}(\vc{r})\big]e^{-i\omega t},\nonumber\\
p^c_1(\vc{r},t)&=&-\omega a\big[\gamma_{lm}p_{lm}(\vc{r})+\gamma_{l+1,m}p_{l+1,m}(\vc{r})\big]e^{-i\omega t},
\end{eqnarray}
with two complex coefficients $\gamma_{lm},\gamma_{l+1,m}$. We allow integer $l\geq 2$ and $m=-l,...,l$. We need two successive coefficients in order to obtain a nonvanishing swimming velocity. Correspondingly we introduce the complex moment vector
\begin{equation}
\label{2.8}\vc{\psi}=(\gamma_{lm},\gamma_{l+1,m}).
\end{equation}
Then the mean second order swimming velocity is given by
\begin{equation}
\label{2.9}\overline{U_2}=\frac{1}{2}\;\omega
a(\vc{\psi}|\du{B}|\vc{\psi}),
\end{equation}
with a dimensionless hermitian $2\times 2$ matrix $\du{B}$. The mean second order rotational swimming velocity is given by
\begin{equation}
\label{2.10}\overline{\Omega_2}=\frac{3}{4}\;\omega
(\vc{\psi}|\du{C}|\vc{\psi}),
\end{equation}
with a dimensionless hermitian $2\times 2$ matrix $\du{C}$. The time-averaged rate of dissipation can be expressed as
\begin{equation}
\label{2.11}\overline{\mathcal{D}_2}=8\pi\eta\omega^2a^3(\vc{\psi}|\du{A}|\vc{\psi}),
\end{equation}
with a dimensionless hermitian matrix $\du{A}$. The matrix elements of the three matrices can be evaluated from Eqs. (2.4)-(2.6).

Explicitly we find for the matrix $\du{A}$
\begin{equation}
\label{2.12}\du{A}_{lm}=\left(\begin{array}{cc}
a_{lm}&0
\\0&a_{l+1,m}
\end{array}\right),
\end{equation}
with element
\begin{equation}
\label{2.13}a_{lm}=a_l\;\frac{(l+m)!}{(l-m)!},\qquad a_l=\frac{(l+1)(2l^2+3l+4)}{l^2(2l+1)}.
\end{equation}
For the matrix $\du{B}$ we find
\begin{equation}
\label{2.14}\du{B}_{lm}=i\left(\begin{array}{cc}
0&b_{lm}
\\-b_{lm}&0
\end{array}\right),
\end{equation}
with element
\begin{equation}
\label{2.15}b_{lm}=b_l\;\frac{(l+m+1)!}{(l-m)!},\qquad b_l=2\frac{(l+2)(2l^2-2l-1)}{l(l+1)(2l+1)(2l+3)},
\end{equation}
and for the matrix $\du{C}$
\begin{equation}
\label{2.16}\du{C}_{lm}=\left(\begin{array}{cc}
c_{lm}&0
\\0&c_{l+1,m}
\end{array}\right),
\end{equation}
with element
\begin{equation}
\label{2.17}c_{lm}=-mc_l\;\frac{(l+m)!}{(l-m)!},\qquad c_l=4\frac{(l+1)(l-2)}{l^2(2l+1)}.
\end{equation}
For $m=0$ the rotational swimming velocity vanishes, but for $m\neq 0$ the swimmer propels itself in helical fashion.

We note that for $m=0$ the expressions for $\overline{U}_2$ and $\overline{\mathcal{D}}_2$ agree with those derived by Felderhof and Jones \cite{3}, as can be seen by putting $\mu_l=-(l-2)\kappa_l/l$ in their expressions (7.8) and (7.15). The expressions agree also with those derived by Blake \cite{2}.

\section{\label{III}Optimal 2-mode helical swimmer}

Optimization of the mean translational swimming velocity for given mean rate of dissipation leads to the eigenvalue problem
\begin{equation}
\label{3.1}\du{B}|\psi_\lambda)=\lambda\du{A}|\psi_\lambda).
\end{equation}
Both matrices $\du{B}$ and $\du{A}$ are hermitian, so that the eigenvalues $\lambda$ are real. For the 2-mode swimmer the eigenvalues are $\pm\lambda_{lm}$ with
\begin{equation}
\label{3.2}\lambda_{lm}=\frac{b_l}{\sqrt{a_la_{l+1}}}\;\sqrt{(l+1)^2-m^2}.
\end{equation}
We denote the corresponding eigenvector as $\vc{\xi}_{lm}$ and normalize such that the first component equals unity,
\begin{equation}
\label{3.3}\vc{\xi}_{lm}=(1,-iy_{lm}),\qquad y_{lm}=\frac{a_{lm}}{b_{lm}}\lambda_{lm}.
\end{equation}
It is evident from Eq. (3.2) that at any $l$ the axial swimmer with $m=0$ is optimal. From Eqs. (2.13) and (2.15) we see that at large $l$ we have $\lambda_{lm}\approx\sqrt{(l+1)^2-m^2}/l$. In Fig. 1 we plot the ratio $R_l=lb_l/\sqrt{a_la_{l+1}}$ as a function of $l$. This shows a monotonic increase with $l$ and a significant deviation from the limit value $R_\infty=1$ at moderate values of $l$. In particular $R_2=0.099$.

The eigenvalue $\lambda_{lm}$ as a function of $(l,m)$ is a measure of the optimal speed at fixed power, since
\begin{equation}
\label{3.4}\lambda_{lm}=\frac{(\vc{\xi}_{lm}|\du{B}_{lm}|\vc{\xi}_{lm})}{(\vc{\xi}_{lm}|\du{A}_{lm}|\vc{\xi}_{lm})}.
\end{equation}
In analogy we define
\begin{equation}
\label{3.5}\rho_{lm}=\frac{(\vc{\xi}_{lm}|\du{C}_{lm}|\vc{\xi}_{lm})}{(\vc{\xi}_{lm}|\du{A}_{lm}|\vc{\xi}_{lm})}.
\end{equation}
This is a measure of the rate of rotation of the optimal swimmer at constant power. We find
\begin{equation}
\label{3.6}\rho_{lm}=-4m \frac{2l^3+2l^2+2l-11}{(2l^2+3l+4)(2l^2+7l+9)}.
\end{equation}
For large $l$ this is approximately $\rho_{lm}\approx -2m/l$.

We denote the reduced power of the optimal swimmer with moments $\vc{\xi}_{lm}$ as $P_{lm}$,
\begin{equation}
\label{3.7}P_{lm}=(\vc{\xi}_{lm}|\du{A}_{lm}|\vc{\xi}_{lm}).
\end{equation}
The time the swimmer needs to move over a distance equal to one diameter is
\begin{equation}
\label{3.8}t_{lm}=\frac{2a}{\overline{U}_2}=\frac{4}{\omega\lambda_{lm}P_{lm}}.
\end{equation}
During this time the swimmer rotates over  the angle
\begin{equation}
\label{3.9}\overline{\Omega}_2t_{lm}=\frac{3\rho_{lm}}{\lambda_{lm}}.
\end{equation}
This increases monotonically with $m$ and is independent of the power. For $m=\alpha l$ the ratio $\rho_{lm}/\lambda_{lm}$ tends to $2\alpha/\sqrt{1-\alpha^2}$ as $l$ tends to infinity.

\section{\label{IV}Many-mode helical swimmer}

 Next we consider swimmers characterized by moment vector $(\gamma_{2m},\gamma_{3m},...,\gamma_{Lm})$ with $L-1$ entries. The corresponding $(L-1)\times(L-1)$-dimensional matrices $\du{A}_{2L,m}$ and $\du{C}_{2L,m}$ are diagonal, and the matrix $\du{B}_{2L,m}$ is tridiagonal. The matrix elements can be read off from Eqs. (2.12)-(2.17). We can simplify the problem by introducing modified moments
 \begin{equation}
\label{4.1}g_{lm}=(-i)^l\sqrt{a_{lm}}\gamma_{lm}.
\end{equation}
With these moments the mean rate of dissipation is
\begin{equation}
\label{4.2}\overline{\mathcal{D}_2}=8\pi\eta\omega^2a^3\sum^L_{l=2}|g_{lm}|^2
=8\pi\eta\omega^2a^3(\vc{g}|\du{A}^{\prime}|\vc{g}),
\end{equation}
where $\du{A}^{\prime}=\du{I}$ with unit matrix
$\du{I}$, and the translational swimming velocity is
\begin{equation}
\label{4.3}\overline{U_2}=-\omega
a\sum^L_{l=2}\lambda_{lm}\mathrm{Re}g_{lm}^*g_{l+1,m}=\frac{1}{2}\omega
a(\vc{g}|\du{B}^{\prime}|\vc{g}),
\end{equation}
where $\du{B}^\prime$ is symmetric with non-zero elements
\begin{equation}
\label{4.4}B^{\prime}_{l,l+1}=B^{\prime}_{l+1,l}=-\lambda_{lm}.
\end{equation}
The eigenvectors of the corresponding eigenvalue problem
\begin{equation}
\label{4.5}
\du{B}^{\prime}|\vc{f}_\lambda)=\lambda|\vc{f}_\lambda),
\end{equation}
can be taken to be real.

The coefficients $\lambda_{lm}$ tend to unity for large $l$ at fixed $m$, so that the problem corresponds to a chain which becomes uniform for large $l$.
The eigenvalue problem Eq. (4.5) for the uniform chain has eigenvalues
\begin{equation}
\label{4.6}
\lambda_q=-2\cos\bigg(\frac{q\pi}{L}\bigg),\qquad q=1,...,L-1,
\end{equation}
and corresponding eigenvectors with components
\begin{equation}
\label{4.7}
f_{k,q}=C_q\sin\bigg(\frac{kq\pi}{L}\bigg),\qquad k,q=1,...,L-1,
\end{equation}
where $C_q$ is a normalization factor. The largest eigenvalue occurs for $q=L-1$. In the limit $L\rightarrow\infty$ the maximum eigenvalue tends to $2$ and the components of the corresponding eigenvector alternate. The observation implies that we can expect the maximum eigenvalue of the actual problem to be slightly less than 2 for large $L$.

We consider in particular $L=10$. Then the maximum eigenvalue is $\lambda_{max}(10,0)=1.158$ for $m=0$. This shows that for moderate values of $L$ the eigenvalue is far less than the limiting value 2. In Fig. 2 we plot the components of the moment vector $\overline{\vc{g}}_0$, normalized to unity, corresponding to $-\lambda_{max}(10,0)$. The components of $\vc{g}_0$ for $\lambda_{max}(10,0)$ when multiplied by $(-1)^l$ are the same. In Fig. 3 we plot the maximum eigenvalue $\lambda_{max}(10,m)$ as a function of $m$. This shows that for small $m$ the helical swimmer is nearly as efficient as the axial swimmer with $m=0$. At $m=9$ the value $\lambda_{max}(10,9)$ equals $\lambda_{9,9}=0.308$ of the 2-mode swimmer with $l=9,\;m=9$.

The helical swimmer with $m\neq 0$ has a nonvanishing rotational swimming velocity given by
\begin{equation}
\label{4.8}\overline{\Omega_2}=\frac{1}{2}\omega
(\vc{g}|\du{C}^{\prime}|\vc{g}),
\end{equation}
where the matrix $\du{C}'$ is diagonal with elements
\begin{equation}
\label{4.9}C'_{ll,m}=-4m\frac{l-2}{2l^2+3l+4}.
\end{equation}
For moment vector $\vc{g}=\varepsilon\vc{g}_0$ with normalization $(\vc{g}_0|\vc{g}_0)=1$ the mean translational swimming velocity is $\overline{U}_2=\frac{1}{2}\varepsilon^2\omega a\lambda_{max}$ and the mean rotational swimming velocity is $\overline{\Omega_2}=\frac{1}{2}\varepsilon^2\omega
(\vc{g}_0|\du{C}^{\prime}|\vc{g}_0)$. In Fig. 4 we plot the matrix element $-(\vc{g}_0|\du{C}^{\prime}|\vc{g}_0)=-(\overline{\vc{g}}_0|\du{C}^{\prime}|\overline{\vc{g}}_0)$ for $L=10$ as a function of $m$. This shows a fairly strong increase with $m$. For $m=9$ the first seven elements of $\overline{\vc{g}}_0$ vanish and the last two elements are $(1/\sqrt{2},1/\sqrt{2})$, in agreement with Eq. (3.3) for the 2-mode swimmer.

In Fig. 5 we show the radial displacement in the meridional plane $
\varphi=0$ at times $t=0,\;T/8,\;T/4$ for the mode given by $\vc{g}_0$ for $L=10,\;m=1$. The maximum eigenvalue is $\lambda_{max}(10,1)=1.150$. It is clear that the surface displacements have the character of a running wave.

Ehlers and Koiller \cite{11} considered a running wave radial displacement proportional to $\cos(n\theta-\omega t)$ with $m=0$ and find the ratio of swimming velocity and power equivalent to $(\psi|\du{B}|\psi)/(\psi|\du{A}|\psi)=\pi/4$ for large $n$. This waveform is appreciably less efficient than the optimum mode. For $m=1$ the authors use a tangential plane approximation method based on Taylor's work \cite{9}. In our scheme the moment vector of the chosen wave $\cos(n\theta+m\varphi-\omega t)$ has an infinite number of components and the calculation of the mean power would need to be carried out till convergence to obtain desired accuracy. It is preferable to base the first order swimming stroke on a solution of the Stokes equations with a finite-dimensional moment vector, as illustrated in Fig. 5.

\section{\label{V}Discussion}

The analysis allows calculation of the mean translational and rotational swimming velocities and the mean rate of dissipation for any prescribed time-periodic radial surface wave on a sphere. It suffices to determine the corresponding moment vector defined as the set of components of the solutions of the Stokes equations defined in Eq. (2.2). The desired quantities are then found as expectation values of matrices with elements given explicitly in Sec. II.

The limitation to radial displacements may be regarded as a constraint resulting from the structure of the swimmer. Elsewhere we have studied swimming with the alternative constraint of tangential surface displacements \cite{6}. Which constraint applies must be determined by detailed consideration of the microorganism. If neither constraint applies one must consider mixed strokes with both radial and tangential components. In that case further analysis is required, as is clear from the axisymmetric case studied by Lighthill \cite{1}, Blake \cite{2}, and Felderhof and Jones \cite{3}.

It would be desirable to extend the investigation to the swimming of a spheroid. The present study may serve as a guide for this more complicated geometry.

\newpage

\section*{Figure captions}

\subsection*{Fig. 1}
Plot of the ratio $R_l=lb_l/\sqrt{a_la_{l+1}}$ as a function of $l$.

\subsection*{Fig. 2}
Plot of the components of the moment vector $\overline{\vc{g}}_0$ for $L=10,\;m=0$ as a function of $l$.

\subsection*{Fig. 3}
Plot of the maximum eigenvalue $\lambda_{max}(10,m)$ as a function of $m$. This characterizes the mean translational swimming velocity
of the optimal many-mode swimmer for $L=10$ at each value of $m$ for given power.

\subsection*{Fig. 4}
Plot of the matrix element $-(\vc{g}_0|\du{C}^{\prime}|\vc{g}_0)=-(\overline{\vc{g}}_0|\du{C}^{\prime}|\overline{\vc{g}}_0)$ for $L=10$ as a function of $m$. This characterizes the rate
of steady rotation of the optimal  many-mode swimmer for $L=10$ at each value of $m$ for given power.

\subsection*{Fig. 5}
Plot of the radial displacement $\xi_r$ in the meridional plane $\varphi=0$ as a function of polar angle $\theta$ at times $t=0$ (solid curve), $T/8$ (long dashes), $T/4$ (short dashes) for the optimal  many-mode swimmer with $L=10,\;m=1$.

\newpage
\clearpage
\newpage
\setlength{\unitlength}{1cm}
\begin{figure}
 \includegraphics{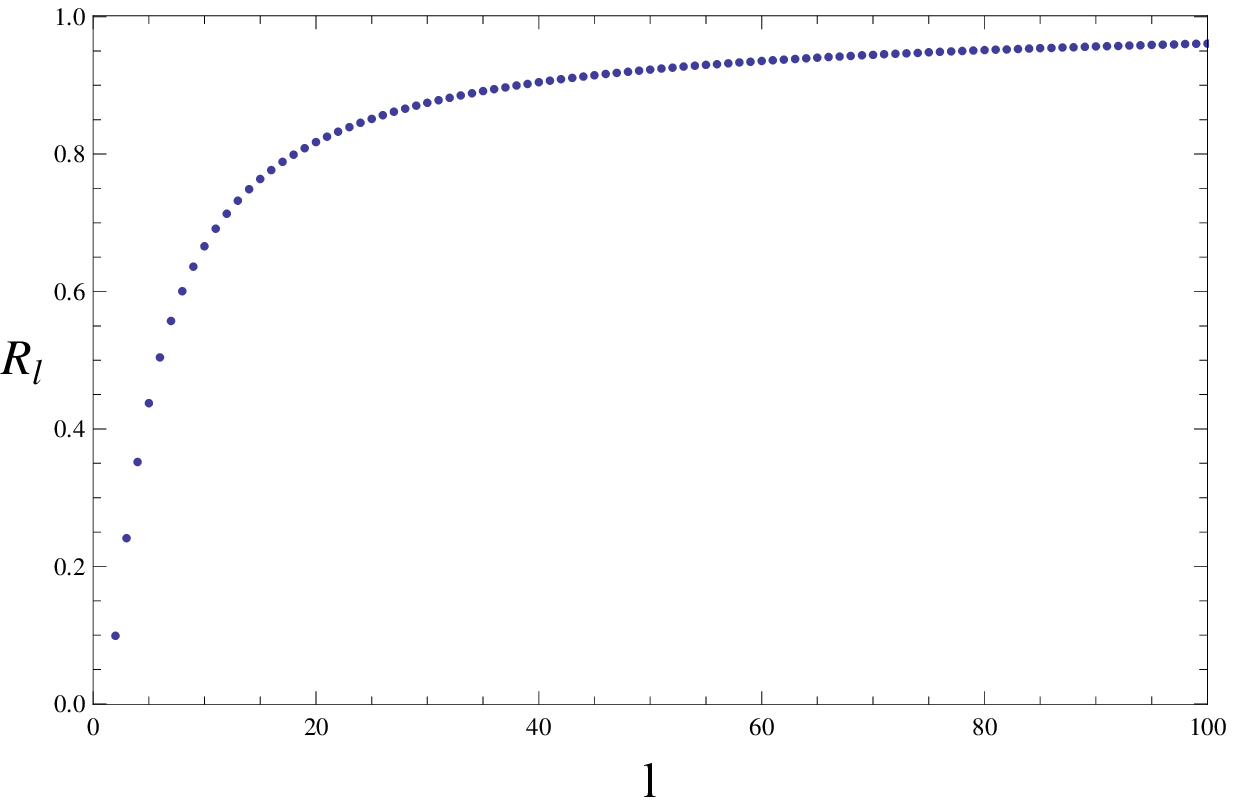}
   \put(-9.1,3.1){}
\put(-1.2,-.2){}
  \caption{}
\end{figure}
\newpage
\newpage
\clearpage
\newpage
\setlength{\unitlength}{1cm}
\begin{figure}
 \includegraphics{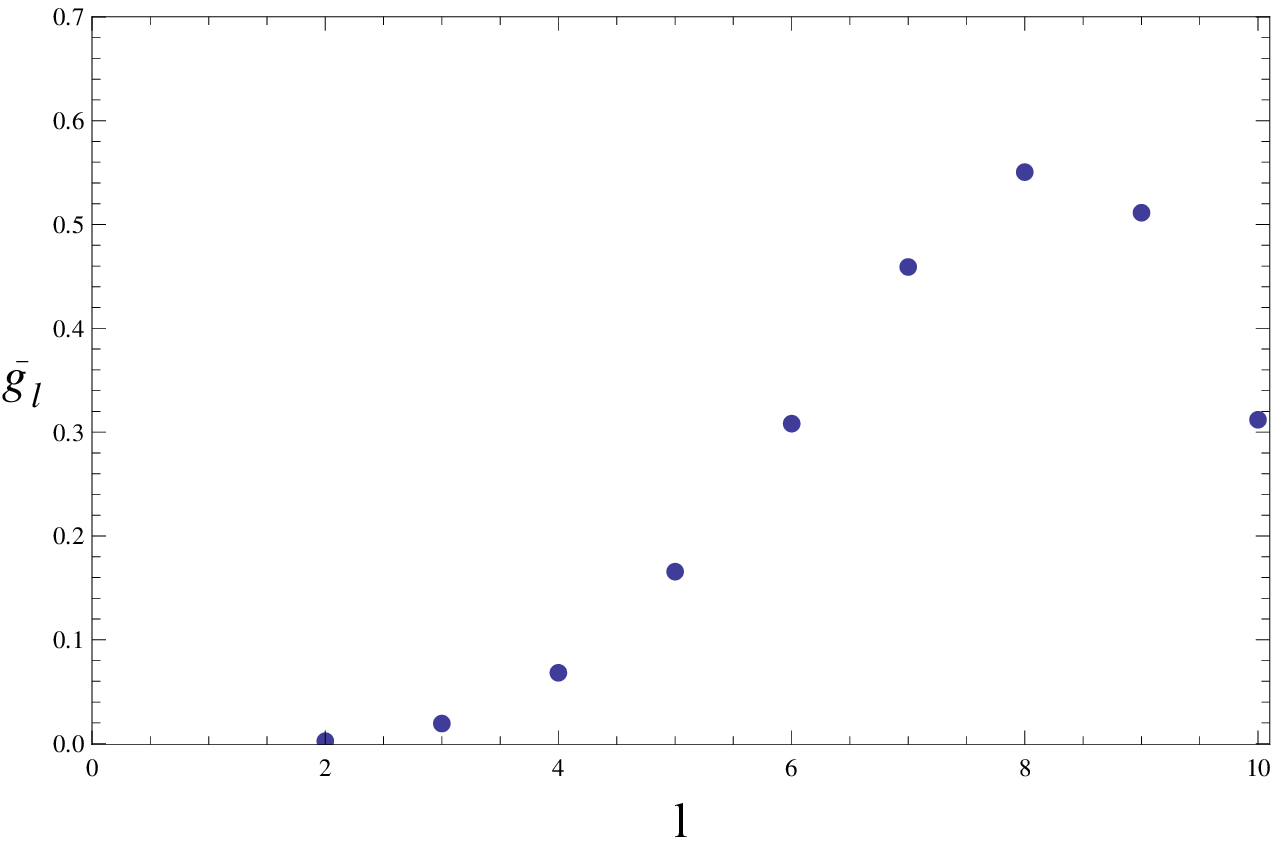}
   \put(-9.1,3.1){}
\put(-1.2,-.2){}
  \caption{}
\end{figure}
\newpage
\clearpage
\newpage
\setlength{\unitlength}{1cm}
\begin{figure}
 \includegraphics{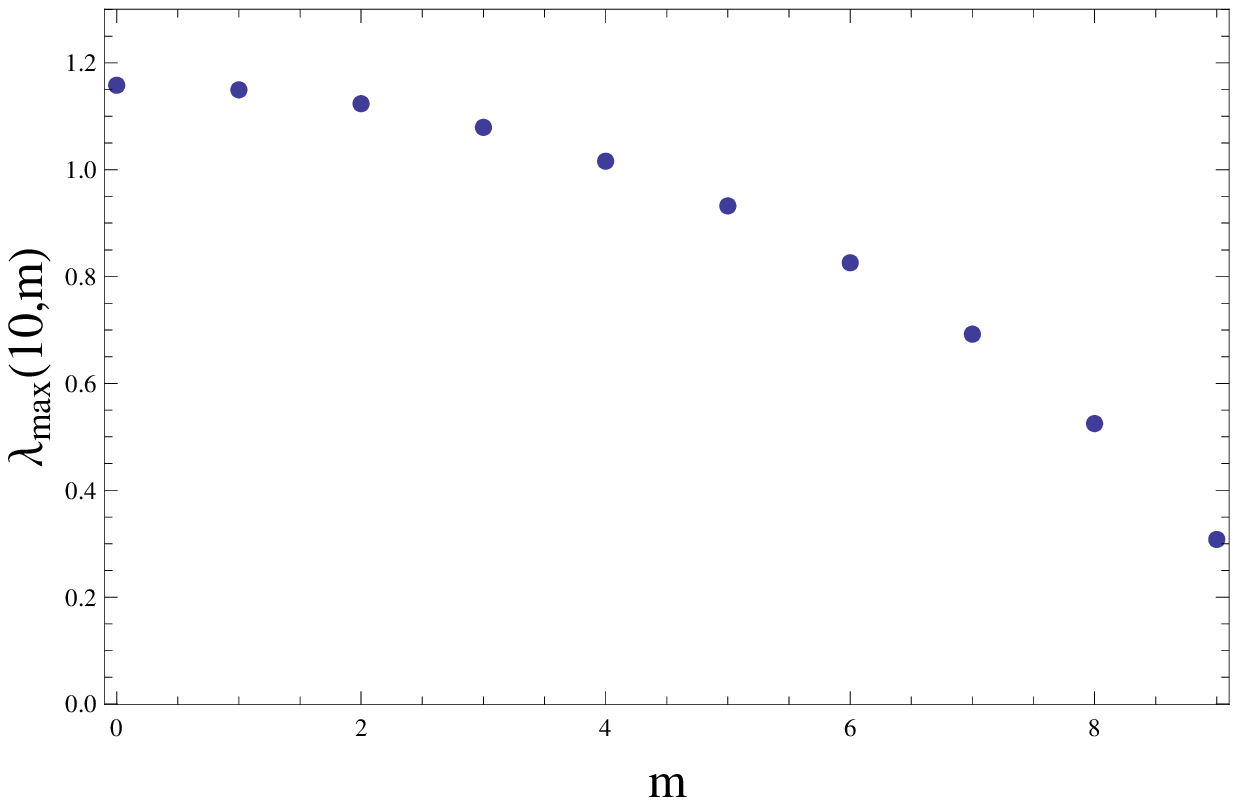}
   \put(-9.1,3.1){}
\put(-1.2,-.2){}
  \caption{}
\end{figure}
\newpage
\clearpage
\newpage
\setlength{\unitlength}{1cm}
\begin{figure}
 \includegraphics{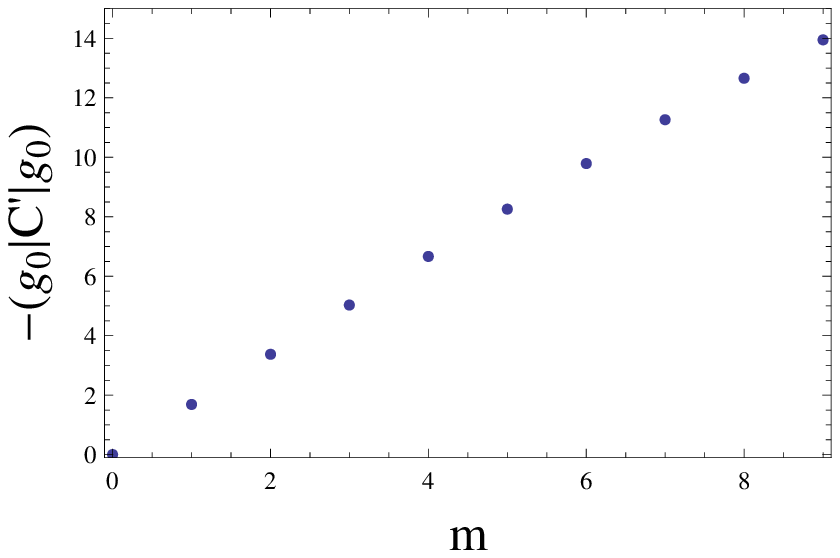}
   \put(-9.1,3.1){}
\put(-1.2,-.2){}
  \caption{}
\end{figure}
\newpage
\clearpage
\newpage
\setlength{\unitlength}{1cm}
\begin{figure}
 \includegraphics{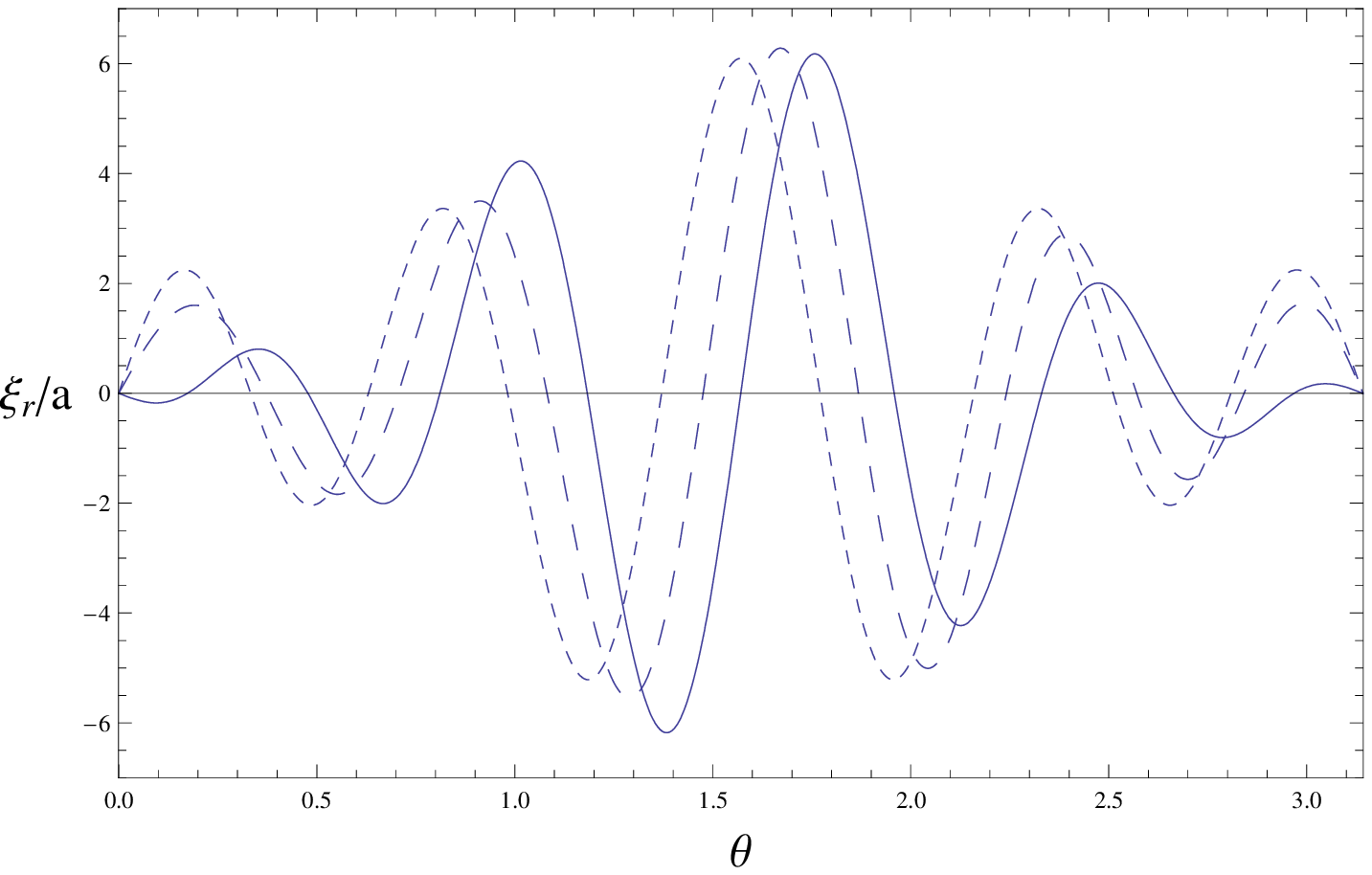}
   \put(-9.1,3.1){}
\put(-1.2,-.2){}
  \caption{}
\end{figure}
\end{document}